# Exfoliation of Few-Layer Black Phosphorus in Low-Boiling-Point Solvents and Its Application in Li-Ion Batteries

Antonio Esau Del Rio Castillo,*,† Vittorio Pellegrini,† Haiyan Sun,† Joka Buha,‡ Duc Anh Dinh, Emanuele Lago,† Alberto Ansaldo,† Andrea Capasso,† Liberato Manna,‡ and Francesco Bonaccorso*,†

†Graphene Labs Istituto Italiano di Tecnologia, via Morego 30, 16163 Genoa, Italy

‡Nanochemistry Department, Istituto Italiano di Tecnologia, via Morego 30, 16163 Genoa, Italy

Corresponding authors: Antonio.delrio@iit.it, Francesco.bonaccorso@iit.it

## Abstract

The liquid-phase exfoliation (LPE) of black phosphorus (BP) is a strategic route for the large-scale production of phosphorene and few-layer BP (FL-BP) flakes. The exploitation of this exfoliated material in cutting-edge technologies, *e.g.*, in flexible electronics and energy storage, is however limited by the fact that the LPE of BP is usually carried out at a high boiling point and in toxic solvents. In fact, the solvent residual is detrimental to device performance in real applications; thus, complete solvent removal is critical. Here, we tackle these issues by exfoliating BP in different low boiling-point solvents. Among these solvents, we find that acetone also provides a high concentration of exfoliated BP, leading to the production of FL-BP flakes with an average lateral size and thickness of ∼30 and ∼7 nm, respectively. The use of acetone to produce less defective few-layer BP flakes (FL-BP$_{acetone}$) from bulk crystals is a straightforward process which enables the rapid preparation of homogeneous thin films thanks to the fast solvent evaporation. The ratio of edge to bulk atoms for the BP flakes here produced, combined with the use of low-boiling-point solvents for the exfoliation process suggests that these thin films are promising anodes for lithium-ion batteries. To this end, we tested Li-ion half cells with FL-BP$_{acetone}$ anodes achieving a reversible specific capacity of 480 mA h g$^{-1}$ at a current density of 100 mA g$^{-1}$, over 100 charge/discharge cycles. Moreover, a reversible specific capacity of 345 mA h g$^{-1}$ is achieved for the FL-BP$_{acetone}$-based anode at high current density (*i.e.*, 1 A g$^{-1}$). These findings indicate that the FL-BP$_{acetone}$-based battery is promising with regards to the design of fast charge/ discharge devices. Overall, the presented process is a step forward toward the fabrication of phosphorene-based devices.

## Introduction

Layered crystals are a class of materials[1,2] that are composed of stacks of atomically thin layers which are held together by van der Waals forces.[3,4] Recently, layered crystals have attracted the attention of the scientific community since their physical and chemical properties change upon thickness reduction; *i.e.*, they change from bulk to single layer.[5,6] These materials are at the origin of several key technological applications,[7–10] including, in particular, those in the energy storage sector. A key example is graphite, which is currently used as a commercial anode material for Li-ion batteries (LIBs) thanks to its stable electrochemical performance.[11] Furthermore, graphene flakes obtained from exfoliated graphite crystals have also shown promising performances, improving at the lab-scale the state-of the-art anode of LIBs thanks to the optimization of the flake aspect ratio.[10,12] Layered crystals beyond graphite are promising materials for ion storage; *e.g.*, some transition metal di-chalcogenides can intercalate ions in their structure and exhibit fast ionic conductivity.[13] Among them, black phosphorus (BP), having a theoretical capacity of 2596 mA h g$^{-1}$,[14–19] which is much higher than those of graphite (372 mA h g$^{-1}$) and graphene (744 mA h g$^{-1}$),[20,21] is a quite promising anode material. Additionally, its charge/discharge rate due to the Li fast diffusion is 10$^4$ times faster than it is in graphene[12] (the diffusion energy barrier in BP is 0.09 eV[17] and in graphene is 0.327 eV[22]). Moreover, when exfoliated into single-layer (phosphorene) or into few-layer BP flakes, the increased available surface area can further enhance the electrochemical activity,[23] which is beneficial for energy storage applications.[14,24] Therefore, exfoliated BP is a promising candidate for the next generation of Li-ion battery anodes.

The exfoliation of bulk BP into single- or few-layer flakes can be performed using micromechanical cleavage (MC),[25,26] which consists of consecutively peeling off crystal layers by using adhesive tape.[27] However, this technique is only suitable for research activities because of both the scalability limitation[28] and morphological heterogeneity of the resulting flakes.[27] In contrast, the liquid-phase exfoliation (LPE)[29–36] is an affordable and scalable alternative to MC.[25,26] The LPE of BP in both aqueous[45] and organic[37–40] solvents has been recently demonstrated, which opened up possibilities to use the exfoliated BP in applications as light absorbers[41,42] and energy storage devices.[14–16,19,43–46] Current approaches for the LPE of BP present several issues especially in aqueous environment, where the chemical integrity of the exfoliated flakes is compromised because of the oxidation promoted by the presence of O$_2$/H$_2$O.[47–49] The formation of phosphorus-oxide species, *i.e.*, P$_2$O$_5$ and P$_2$O$_4$,[50] has been reported when the BP is exposed to air. These processes increase the roughness of the flakes and accelerate their degradation.[51] Performing the LPE in pure organic solvents solves this issue, since the presence of water and O$_2$ can be avoided. The organic solvents that are commonly used to exfoliate BP are generally toxic (health code ≥2 NFPA704),[52] and have boiling points (bp) usually above 100 °C, *e.g.*, N-methyl-2-pyrrolidone (NMP, bp = 202 °C),[37] N,N-dimethylformamide (DMF, bp = 153 °C),[38,53] N-cyclohexyl-2-pyrrolidone (CHP, bp = 284 °C),[39] or formamide (bp 210 °C).[54] The solvent bp is critical for several applications because the solvent removal is of paramount importance for the realization of high-performance anodes of batteries,[55–57] and it is also relevant for the development of electronic[58,59] and optoelectronic devices.[42,60] In general, the solvent removal is performed by heating the deposited sample or device above the solvent bp. However, annealing procedures always run the risk of either degrading the material or damaging the device. Furthermore, in some cases the solvent degrades when heated, *i.e.*, NMP;[61] thus it leaves contaminants or impurities on the as prepared devices,[62] which are detrimental to their performance.[63] A possible solution for these problems is to use an easy to-remove solvent, preferably one that is not toxic, not degrading with the annealing temperature. To date, however, a clear solution to overcome such an issue has not been found yet.

Driven by this need, we exfoliated BP by LPE using 14 different solvents, which were selected after having considered the different values of surface tension (γ), bp, and/or Hansen and Hildebrand solubility parameters. The analysis allows us to determine the surface energy and Hansen and Hildebrand parameters of BP, which are relevant for the selection of the ideal solvent for the LPE of BP. In particular, the selected solvent should be able to (i) exfoliate BP; (ii) keep a stable dispersion of the exfoliated flakes (*i.e.*, the exfoliated flakes should not flocculate or precipitate); (iii) prevent the degradation of the exfoliated flakes by oxidation; and (iv) be easily removed without leaving impurities.

Among the 14 solvents tested, our results (optical extinction spectroscopy) confirm the exfoliation of BP in CHP, NMP, DMF, diethylcarbonate, 3-chloroethylene, and acetonitrile. Our data indicate that CHP is the solvent that promotes the highest concentration compared with all the other solvents. Interestingly, the exfoliation is also possible in acetone, a well-known nontoxic solvent with a low boiling point.[52] The exfoliation in acetone is attractive for real applications, *e.g.*, polymer composites and functional inks, for both of which the drying time and toxicity are key factors to be considered. For these reasons, we provide here an extensive analysis of exfoliated BP flakes in acetone, addressing the morphological properties, *i.e.*, flake sizes and chemical structure, comparing the obtained results with the one using the state-of-the-art solvent[39] such as CHP.

The morphological and structural characterization reveals that the exfoliated flakes in acetone are undamaged by the LPE process, and have an average lateral size of ∼30 nm and a thickness of ∼7 nm. Electron energy loss spectroscopy (EELS) shows that aging (oxidation) the exfoliated BP flakes in acetone is comparable with the ones processed in CHP, for which the ratio of the oxidized against non-oxidized phosphorus ($P_xO_y/P_0$) is 1:4 after 2 weeks of aging. Finally, we demonstrate the feasibility and up-scalability of our approach by designing homogeneous films of few-layer BP (FL-BP) flakes exfoliated in acetone and used as anodes for lithium-ion batteries.

The as-produced anodes, based on the BP flakes that were exfoliated in acetone, demonstrated a reversible specific capacity of 480 mA h g$^{-1}$ after 100 cycles at a current density of 100 mA g$^{-1}$, with a Coulombic efficiency (CE) > 99.5%. These electrochemical performances exceed the ones obtained with the anodes that were produced by the BP flakes exfoliated in CHP (which had a specific capacity of 185 mA h g$^{-1}$ after 100 cycles at a current density of 100 mA g$^{-1}$, with a Coulombic efficiency of 99.4%).

# Experimental section

## Preparation of the Dispersions.

Black phosphorus (500 mg, from Smart Elements) is pulverized with a mortar and pestle. The selected solvents (*i.e.*, acetone, toluene, chloroform, 2-propanol, trichloroethylene, methanol, ethylene glycol, acetonitrile, ethanol, *n*-hexane, NMP, CHP, DMF, and diethyl carbonate) have different γ, bp, and/or Hansen solubility parameters (see the Appendix). All the solvents are of anhydrous grade and were purchased from Sigma-Aldrich.

For the analysis of the dispersibility and stability of exfoliated BP, 20 mg of pulverized BP and 20 mL of solvents are mixed using a sonic bath (VWR, USC2600THD) for 6 h, followed by centrifugation at 900g for 60 min to promote the precipitation of the thicker and un-exfoliated flakes. The precipitation of thick or un-exfoliated flakes promotes the enrichment of BP flakes in dispersion with a specific lateral size and thickness. The size selection depends on the solvent density and viscosity as well as on the applied centrifugal force, which is discussed in the Results and Discussion section. The centrifugation is carried out in a Sigma 2-16K centrifuge (11170-bucket 2 × 13299 rotor). After the centrifugation, the supernatant is collected and subjected to another centrifugation run at 900g for 30 min to further purify the BP dispersions. The pulverization and weighting of the BP crystals and the balancing and sealing of the centrifuge tubes are carried out in a nitrogen-filled glovebox. For the aging analysis, the samples are stored for three months at room temperature in a closed transparent glass vial. It is noted that, after the LPE process, all the other experimental processes for the material production are conducted outside the glovebox.

## Characterization of the Dispersions. Optical Extinction Spectroscopy.

The optical extinction spectroscopy is carried out by a Cary Varian 5000 UV–vis. For measurement of the extinction spectra, FL-BP dispersions in the different solvents are diluted 1:25 with the respective pure solvents. The dispersions in acetone are diluted at different ratios to determine the extinction coefficient of FL-BP. The dilutions prepared are 1:1, 1:2, 1:5, 1:10, 1:20, 1:50, and 1:100. For each sample, the extinction spectra (absorbed plus scattered light) of their corresponding pure solvents are subtracted from the sample spectrum. After the extinction measurement, the samples that are dispersed in acetone are dried, and the residue powder is weighted. The initial volume of the solvent and the mass of the dried powder gave the exact concentration for each dilution. The optical extinction coefficient is determined by using the Beer–Lambert law ($E = \alpha C_{ph} l$, in which $E$ is the optical extinction at 600 nm, $\alpha$ is the extinction coefficient, $C_{ph}$ is the concentration of the exfoliated BP, and $l$ is the path length, 0.01 m).

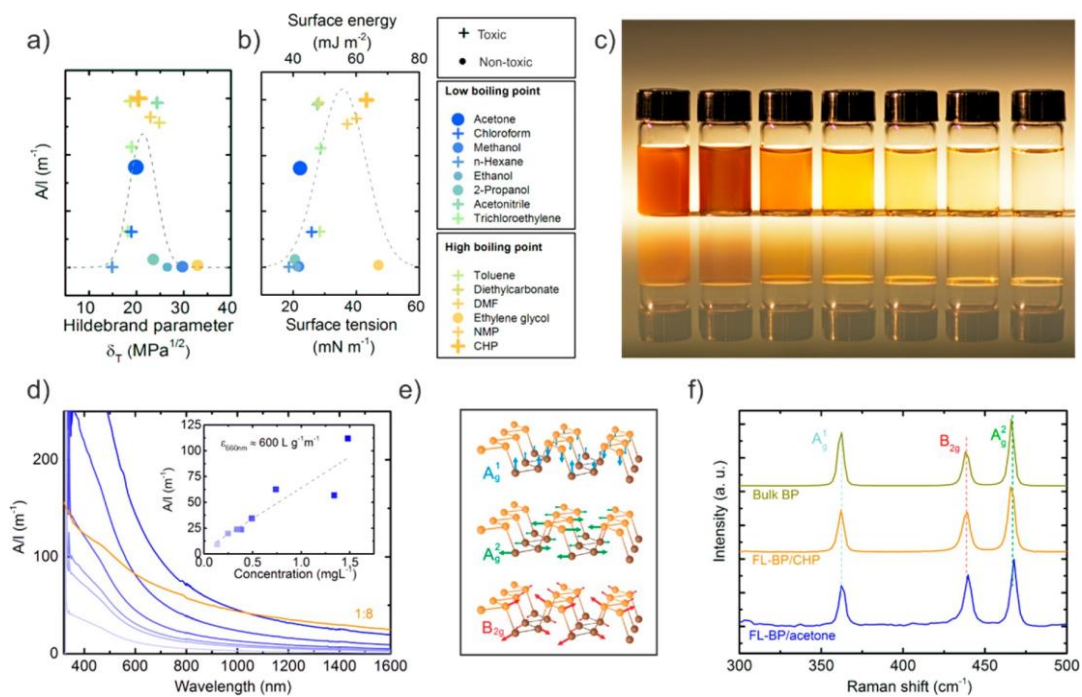

*Figure 1*. Extinction coefficient of BP dispersed in different solvents after the LPE process, plotted as a function of (a) Hildebrand parameter, and (b) surface tension (lower axis) and surface energy (upper axis). Dots represent the low/nontoxic solvents (health code ≤1 NFPA704); crosses denote the highly toxic solvents (health code ≥2 NFPA704). Dots and crosses with colours from blue to light-green represent the solvents with bp < 100 °C, while the ones from turquoise to orange represent solvents with bp > 100 °C. (c) Photograph of the FL-BP dispersions in CHP (first from the left) and acetone at different concentrations (from the second left to right). (d) Extinction spectra of the FL-BP dispersions in acetone at different concentrations, in different blue tones, and FL-BP dispersions in CHP, in orange. The inset shows the calibration curve according to the Beer–Lambert law, yielding an extinction coefficient of ∼600 L g−1 m−1, at 660 nm. (e) Scheme of the FL-BP Raman active modes. (f) Raman spectra of the bulk BP (dark yellow line), of the FL-BP cast from a CHP dispersion (orange line), and of FL-BP cast from a dispersion in acetone (blue line).

### Transmission Electron Microscopy.

The BP dispersions in acetone and CHP are diluted 1:10 with their corresponding pure solvents. A 100 µL portion of the resulting dilutions are drop-cast at room temperature onto copper holey carbon grids (300 mesh) and subsequently dried under vacuum overnight. TEM images are taken with a JEOL JEM 1011 transmission electron microscope operating at 100 kV. The high-resolution TEM (HRTEM), scanning TEM (STEM), energy-dispersive X-ray spectrometry (EDS), and energy filtered (EFTEM) elemental mapping are carried out on a JEOL JEM 2200FS microscope operating at 200 kV and equipped with a field emission gun, CEOS image Cs corrector, Bruker Quantax EDS system with an XFlash 6T-60 silicon drift detector and an in-column Omega filter. The electron energy loss spectroscopy (EELS) data are collected in TEM mode using an FEI Tecnai F20 TEM operating at 200 kV and equipped with a Gatan Enfinum SE spectrometer. The EELS spectra are acquired with a collection semiangle of 100 mrad.

### Raman Spectroscopy.

For the Raman measurements, the BP dispersions are diluted 1:10 with the pure solvents, *i.e.*, with both acetone and CHP, and drop-cast onto a Si wafer (LDB Technologies, Ltd.) with 300 nm of thermally grown $SiO_2$. The measurements are carried out with a Renishaw inVia spectrometer using a 50× objective, an excitation wavelength of 514.5 nm, and an incident power of ∼1 mW on the samples.

### Electrode Preparation.

The copper foil (25 μm think, 99.98% trace metals basis, Sigma-Aldrich) is cut into round disks with a diameter of 0.015 m and cleaned with acetone (Sigma-Aldrich) in an ultrasonic bath for 10 min. Then, the copper disks are dried for 2 h in a glass vacuum oven (BÜCHI, B-585) at 80 °C with $10^{-3}$ bar pressure. The FL-BP dispersions in CHP and acetone are mixed with carbon black (CB, super-P, Alfa-Aesar) and polyvinylidene fluoride (PVdF, Sigma-Aldrich), with a mass ratio of 3:1:1, then drop-cast onto the cleaned copper substrates inside the glovebox at room temperature. The resulting FL-BP-coated electrodes are subsequently dried in a vacuum oven at 120 °C for acetone, and to 180 °C for the CHPdispersed sample, at $10^{-3}$ bar pressure for 30 min. The average mass loading of FL-BP (not including CB or PVdF) that is obtained from both the CHP and acetone dispersions is 0.6 mg $cm^{-2}$ for both samples.

### Electrode Characterization.

Scanning Electron Microscopy (SEM). The battery electrodes (exfoliated BP flakes deposited onto Cu substrates) are imaged using a field-emission scanning electron microscope FE-SEM (JEOL JSM-7500 FA). The acceleration voltage is set to 5 kV. The collected images are acquired using the in-lens sensor (secondary electron in-lens image).

### Electrochemical Characterization.

The FL-BP-coated electrodes are then assembled in coin cells (2032, MTI) in half-cell configuration using a 1 M $LiPF_6$ in ethylene carbonate/dimethylcarbonate (EC/ DMC, 1:1 volume ratio) mixed solvent electrolyte (LP30, BASF), a glass fibre separator (Whatman GF/D), and a lithium foil (SigmaAldrich) circular electrode. The coin-cell assembly is carried out in an argon-filled glovebox ($O_2$ and $H_2O$ < 0.1 ppm) at 25 °C. The anodes are tested against the Li foil for constant current charge/discharge galvanostatic cycles, performed at different specific currents using a battery analyser (MTI, BST8-WA). They are also tested for cyclic voltammetry (CV) at a scan rate of 30 μV $s^{-1}$ using a battery tester (Biologic, MPG2). All the electrochemical measurements are performed at room temperature.

## Results and discussion.
### Solvent Analysis.

For exfoliation and stabilization of BP in a solvent, the Gibbs free energy of the mixture solvent/layered material must be minimized.[29,36,64,65] This condition can be endorsed if the γ of the solvent is equivalent to the surface free energy of the material:[29]

$$\gamma = E_{surface\,solvent} - TS_{surface\,solvent} \quad (1)$$

in which E is the solvent surface energy, T is the absolute temperature, and S is the solvent surface entropy (which generally takes a value of $10^{-3}$ J m$^{-2}$ K$^{-1}$[29,66,67]). Moreover, the matching of the Hansen or Hildebrand parameters of the solvent with the ones of the layered material facilitates the exfoliation process.[66–69] The Hildebrand parameter ($\delta_T$) is widely used in polymer science, and is defined as the square root of the cohesive energy density:[65,68]

$$\delta_T = \sqrt{\frac{\Delta H_v - RT}{V_m}} \quad (2)$$

in which $\Delta H_v$ is the enthalpy of vaporization, $R$ is the ideal gas constant, and $V_m$ is the molar volume. The Hildebrand parameter is used to evaluate the solubility or "dispersibility" of a material in a known solvent.[69,70] However, in some specific cases, the Hildebrand parameter is not sufficient to describe and evaluate the dispersibility of a material in a solvent. For example, the Hildebrand parameter of graphene is ~23 MPa$^{1/2}$,[71] according to the solubility theory. A solvent with this $\delta_T$ value, e.g., 2-propanol with $\delta_T \approx 23.8$,[65] should form a stable dispersion of graphene which, however, has not been experimentally demonstrated. The reason lies in the fact that the Hildebrand parameter does not consider the hydrogen bonding and polar interactions.[65] In contrast, the Hansen solubility parameter splits the cohesive energy ($\delta^2_T$) into three components: the polar contribution ($\delta_p$), the dispersive component ($\delta_d$), and the hydrogen-bonding ($\delta_h$):[65]

$$\delta_T^2 = \delta_d^2 + \delta_p^2 + \delta_h^2 \quad (3)$$

The γ, Hildebrand and Hansen parameters of the majority of solvents are reported in the literature.[65] In contrast, the surface energy and the Hildebrand and Hansen parameters of the materials that are under consideration need an experimental estimation. A common way to obtain these data is to disperse the material in different solvents with a known γ, and with known Hildebrand and Hansen parameters. The dispersed material is quantified either directly by evaporating the solvent and weighting the solid fraction, or indirectly by measuring the optical extinction of the material dispersed in the supernatant. Finally, when the solvent parameter value, e.g., the Hildebrand parameter, is plotted against the optical extinction, the maximum of the data distribution indicates the Hildebrand parameter of the dispersed material. The same analysis can be performed to estimate the Hildebrand parameters or to obtain the surface energy (see eq 1).

Following this approach, we tested the exfoliation of BP in different solvents, most of which had been previously used for the LPE of other layered crystals.[29,36] Subsequently, by using the known values of γ and the Hansen and Hildebrand solubility parameters of the solvents as well as the optical extinction of BP dispersed on each solvent, we were able to estimate the surface energy, and the Hansen and Hildebrand parameters of the exfoliated BP flakes. Figure 1a,b shows the solvent/BP dispersibility analysis in terms of the Hildebrand parameter and γ, respectively. The dot distribution in Figure 1 fits a curve that peaks at a value close to 21 MPa$^{1/2}$. This value indicates the Hildebrand parameter of the BP,[65,68,69] and it is in agreement with the one previously reported.[72] The data distribution in Figure 1b shows that the solvents that are able to exfoliate bulk BP have a γ in the range 25–40 mN m$^{-1}$. With the application of eq 1, these values give a BP surface energy in the range 50–65 mJ m$^{-2}$. The Hansen parameters of BP, shown in the Appendix, provide $δ_p$, $δ_h$, and $δ_d$ values in the range 5–12, 5–10, and 15–18 MPa$^{1/2}$, respectively.

Of the 14 tested solvents, optical extinction spectroscopy (see Figure S1 in the Appendix) indicates that 7 are able to exfoliate/ disperse BP: CHP, NMP, DMF, diethyl-carbonate, acetonitrile, trichloroethylene, and acetone.

In particular, three of these solvents have a low boiling point (<100 °C), *i.e.*, trichloroethylene, acetonitrile (health code 2 NFPA704),[52] and acetone, the latter of which is the only nontoxic solvent (health code ≤1 NFPA704).[52] Thus, in the following part of the Article, we fully characterize the exfoliated flakes in acetone and in CHP, the latter being used as a reference solvent.

Figure 1c shows vials with the BP exfoliated in CHP (vial on the left) and the BP exfoliated in acetone at different remove ratios (from the second left to right: no dilution, 40%, 30%, 20%, 10%, and 5%). Their corresponding extinction spectra are reported in Figure 1d, with the extinction coefficient for BP flakes dispersed in acetone being shown as an inset. The slope of this curve indicates that the extinction coefficient is 600 g L$_{-1}$ m$_{-1}$.

There is discrepancy between the extinction coefficient measured in this work and the previous values reported in the literature (see Table 1). This difference is due to the diverse particle size distributions(thickness/lateral size), the refraction indexes of the solvents, and the wavelength

*Table 1. Reported Extinction Coefficients for Liquid-Phase Exfoliated FL-BP in Diverse Solvents*

| Solvent | Thickness (nm) | Lateral size (nm) | Wavelength (nm) | Extinction coefficient (L g$_{-1}$ m$_{-1}$) | Ref. |
|---|---|---|---|---|---|
| DMF | 10 | 200[a] | 1176 | 4819 | 53 |
| DMSO | 20 | 400[a] | 1176 | 5373 | 53 |
| NMP | 10 | 100 | 660 | 263 | 74 |
| H$_2$O | 5 | 100 | 660 | 209 | 75 |
| CHP | 6 | 100 | 465 | 1500[b] | 39 |
| Acetone | 5 | 30 | 660 | 600 | this work |

[a] Lateral size estimated by dynamic light scattering. [b] Value of absorption coefficient

at which the measurement is carried out.[73] The concentration of the FL-BP flakes in CHP is obtained using $\varepsilon_{465nm}$ = 1500 L g$^{-1}$ m$^{-1}$,[39] attaining 0.6 g L$^{-1}$, and the concentration of BP flakes in acetone is obtained with the estimated $\varepsilon_{660nm}$ = 600 L g$^{-1}$ m$^{-1}$, indicating a concentration of 0.35 g L$^{-1}$.

## Morphological Characterization.

Raman spectroscopy gives important information about the vibrational modes of exfoliated crystals. The Raman spectra of BP consist of three peaks, one out-of-plane mode ($A^1_g$, located at 365 cm$^{-1}$) and two in-plane modes ($A^2_g$ and $B_{2g}$, located at 471 and 440 cm$^{-1}$ respectively, Figure 1f).[76,77] The positions and the intensity ratios between these peaks change depending on the T,[78] oxidation,[47,79] strain,[80,81] and number of layers.[38,76] Figure 1f shows the Raman spectra of the starting bulk material (dark yellow), the BP exfoliated in CHP (orange), and the one in acetone (blue). The exfoliated samples display $A^1_g$ at ~362.0 cm$^{-1}$, $B_{2g}$ at ~434 cm$^{-1}$ and $A^2_g$ at ~467 cm$^{-1}$, which is consistent with previous studies on liquid-phase exfoliated BP, obtaining FL-BP.[39,53,74,75] The Raman spectra on both samples, compared with the one of bulk starting material, (Figure 2f) suggest that the LPE process does not damage the BP structure, and the number of layers is reduced with respect to bulk BP.[47]

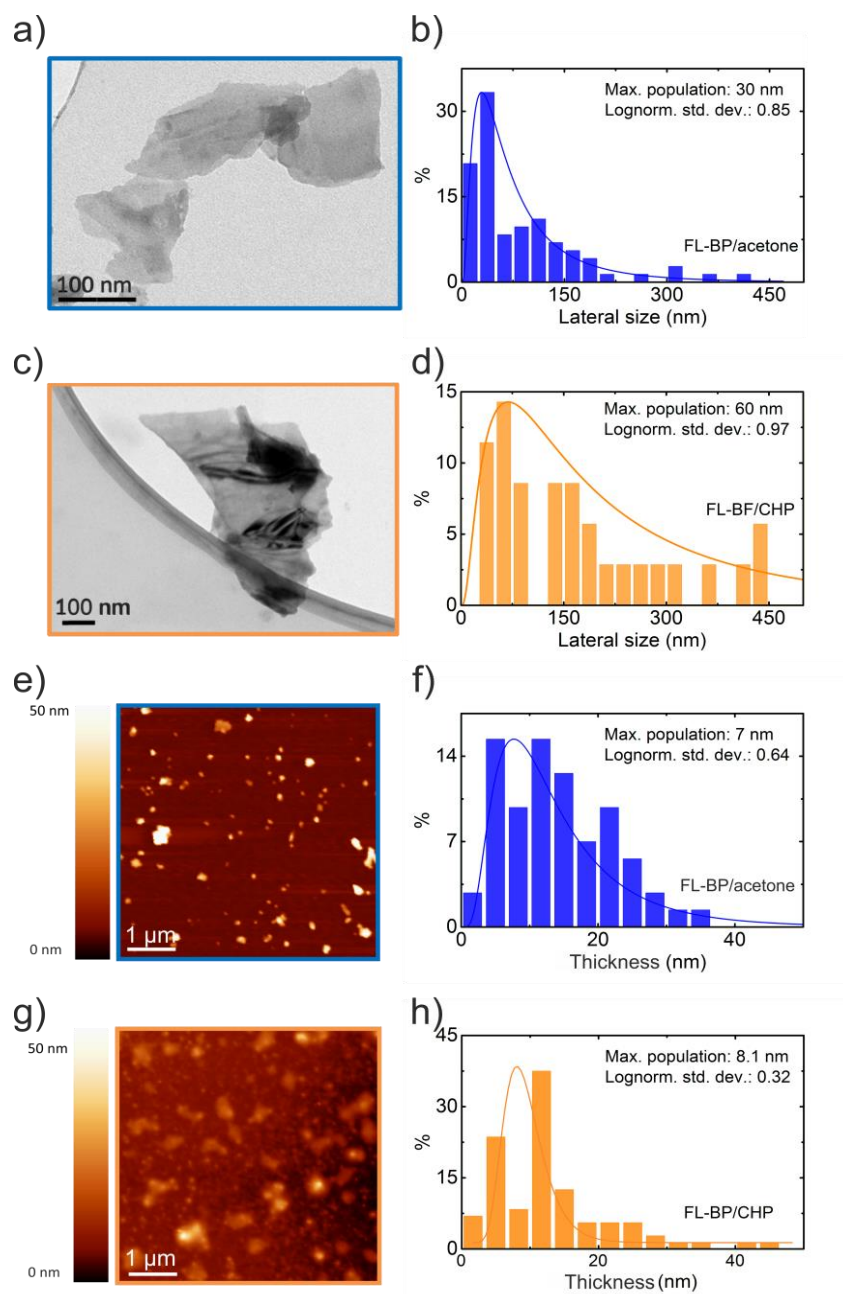

*Figure 2. Transmission electron microscopy of FL-BP in (a) acetone and (c) CHP, and (b, d) their corresponding lateral size distributions. Atomic force microscopy images of (e) FL-BPacetone and (g) FL-BPCHP, and (f, h) their corresponding thickness distributions.*

The TEM characterization provides detailed insight into the morphology and structure of the exfoliated BP flakes. The lateral size distribution analysis indicates that the FL-BP flakes exfoliated in acetone are smaller, with a lateral size of 30 nm (log-normal standard deviation[82] 0.85, Figure 2a,b), compared with the ones exfoliated in CHP, which are 60 nm (log-normal standard deviation 0.97, Figure 2c,d). It is important to analyse the lateral size/thickness differences between the exfoliated BP flakes that are obtained in CHP and acetone. The size selection is carried out during a purification step by step means of sedimentation-based separation (SBS).[28,30,83] The SBS is generally applied for the sorting by size of (nano)particles/ (nano)materials dispersed in a medium subjected to a force field.[84] The sum of the forces acting on the dispersed flakes during centrifugation is[84]

$$F_c - F_b - F_f = F_{tot} \qquad (4)$$

in which the centrifugal force is $F_c = m_p\omega^2 r$, (proportional to the mass of the flake itself, $m_p$; to the distance from the rotational axes, r; and to the square of the angular velocity, $\omega$). The buoyant force is $F_b = -m_s\omega^2 r$, (in which $m_s$ is the mass of the displaced solvent), and the frictional force is $F_f = -fv$, (proportional to the force acting on the particles while moving with a sedimentation velocity, v, in a fluid times the friction coefficient, f, between the solvent and the flake itself, being $f = 6\pi\eta R_e F_t$, for an oblate shape flake; here, $R_e$ is the equivalent radio of the flake being considered as an ellipsoid and $F_t$ the frictional ratio). The volume of the particle can be calculated as its mass ($m_p$) divided by its density ($\rho_p$); the mass of the displaced solvent is $m_s = m_p\rho_s/\rho_p$, in which $\rho_s$ is the density of the solvent;[85] considering that $F_{tot} = 0$ at equilibrium, it is possible to know the sedimentation velocity (v), or the sedimentation rate (S):

$$v = S = \frac{m_p\left(1 - \frac{\rho_s}{\rho_p}\right)}{f} \qquad (5)$$

See the Appendix for the detailed derivation of S. From eq 5 we obtain an S value of 4.76 × 10$^{-13}$ s for FL-BP$_{CHP}$, and 8.34 × 10$^{-12}$ s for FL-BP$_{acetone}$. This result indicates that the flakes of the same size that are dispersed in acetone will precipitate 42 times faster than the ones dispersed in CHP (see Table S2). Table S3 shows the acceleration needed to selectively sort BP flake sizes. The FL-BP flake thickness distribution, estimated by AFM analysis, indicates that the BP flakes exfoliated in acetone peak at 7 nm (~13 staked phosphorene layers, Figure 2e,f), while the ones in CHP peak at 8.1 nm (~16 staked phosphorene layers, Figure 2g,h). The TEM and AFM analysis indicates that FL-BP can be produced using either CHP or acetone.

Figure 3a,f reports the STEM images of FL-BP produced in acetone and CHP (FL-BP$_{acetone}$ and FL-BP$_{CHP}$, respectively). The compositional mapping by EDS (Figure 3b–d,g–i) shows that flakes are composed of P with no appreciable presence of O. Electron energy loss spectroscopy analysis is performed to determine the chemical bonding in the sample, in particular the possible oxide formation during the exfoliation process. These data are presented in Figure S3 of the Appendix. The EELS spectra of the pristine BP sample are characterized by a main sharp P–L$_{2,3}$ edge at 130.2 eV.[86] The peak at 136 eV is assigned to the P–L$_{2,3}$ edge when P forms an oxide (P$_2$O$_5$).[87]

The spectra from the exfoliated samples (Figure S4) exhibit only a single clear P–L$_{2,3}$ edge at 130 eV, and there is no indication of oxidation, which is in agreement with the elemental EDS mapping reported in Figure 3d,i for FL-BP$_{acetone}$ and FL-BP$_{CHP}$ samples, respectively. Moreover, after the FL-BP$_{acetone}$ and FL-BP$_{CHP}$ production, both samples retained the crystal structure of FL-BP, as shown by HRTEM images in Figure 3e,j.

## Aging Study.

The structural analysis of the exfoliated flakes is a challenging task due to the crystalline degradation of the thinnest flakes upon exposure to ambient conditions.[47] The degradation of FL-BP flakes is due to the presence of oxygen groups favoring the formation of PO groups.[39,50,54,74,88] In the case of LPE–BP, it has been reported that CHP and NMP form solvation shells adjacent to the BP surface, which prevents oxidation.[39] In light of this, we performed an aging study comparing FL-BP$_{acetone}$ with FL-BP$_{CHP}$. We used high resolution and analytical TEM techniques, in particular EELS, at different storage times during a three month period. The comparison of the EELS spectra collected from the flakes after different aging times (see the Appendix, Figures S3 and S4) indicates that both samples undergo gradual oxidation after the exfoliation. Detailed characterization, however, shows that this process does not result in a significant structural degradation of the flakes within the first 2 weeks of storage in the respective solvents.

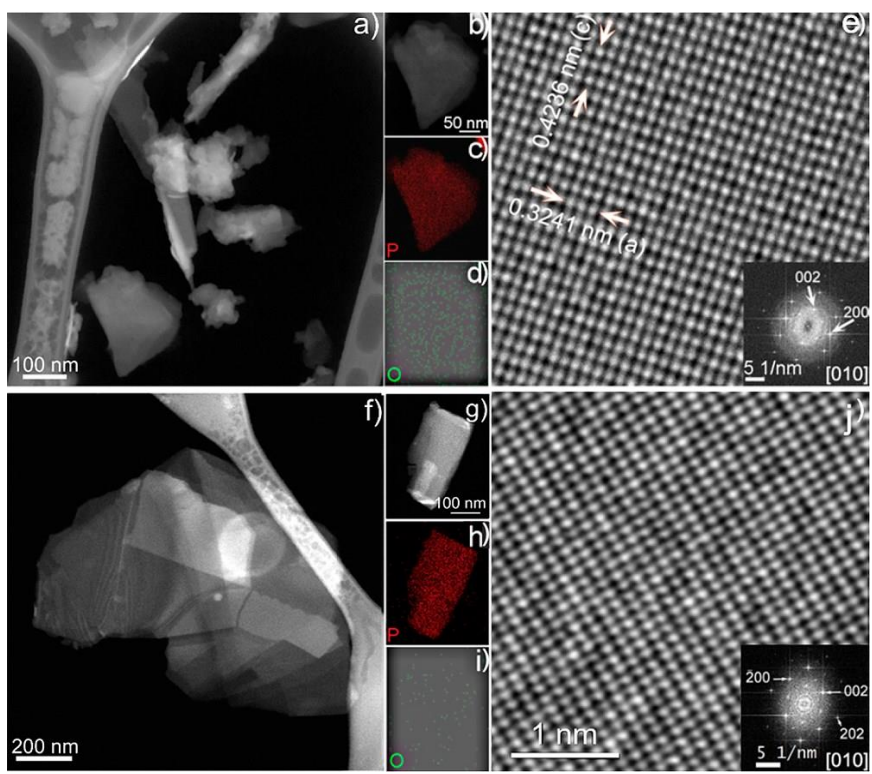

*Figure 3*. (a) STEM image of FL-BPacetone sheets. (b) STEM image of a selected FL-BPacetone sheet and (d) corresponding EDS maps showing the distribution of (c) P and (d) O. (e) HRTEM image of an FL-BPacetone sheet oriented in its 010 axis with the FT reported as an inset. (f) STEM image of several overlapping FL-BPCHP sheets. (g) STEM image of an FL-BPCHP sheet with the EDS maps showing the distribution of (h) P and (i) O. (j) HRTEM of the FL-BPCHP sheet in the 010 orientation with the FT as an inset.

After 12 weeks of aging time, both FL-BP$_{acetone}$ and FLBP$_{CHP}$ samples retain their crystallinity, as is evident from the HRTEM image of an FL-BP$_{acetone}$ flake, shown in the 110 zone axis (Figure 4a). However, the degradation of the flakes is visible as a surface-localized amorphous coating which is 2–5 nm thick, as can be appreciated from the side-view image of the flake reported in Figure 4b. The distance between two adjacent BP layers is 0.52 nm,[89] as is shown by the arrows (see Figure 4b). The compositional analysis of this flake (Figure 4c,d) by EFTEM indicates that the amorphous surface layer is oxygen-rich (Figure 4c), while the central crystalline part of the flake remains phosphorus-rich (Figure 4d). The presence of a thick amorphous layer is also evident on the FL-BP$_{CHP}$ (see arrows in Figure 4e). Fragmentation accompanying oxidation is also observed with thin flakes, such as the one shown in Figure 4f,g, overlapping the thicker (darker in contrast) flake. The degradation of the thinner flake (light contrast) is clearly visible from the EFTEM map of phosphorus (see arrows in Figure 4h).

The spectroscopic and morphological characterizations show that the exfoliation of bulk BP can be performed in acetone as successfully as in high-boiling-point solvents such as CHP. Detailed structural and compositional characterization of the exfoliated material also demonstrates that the oxidation by aging FL-BP$_{acetone}$ is similar to aging FL-BP$_{CHP}$. The exfoliation of BP in acetone is therefore an affordable alternative to the exfoliation process carried out exploiting high-boiling-point and toxic solvents, offering a safe and sustainable route for the exfoliation, storage, and deposition processes of FL-BP flakes.

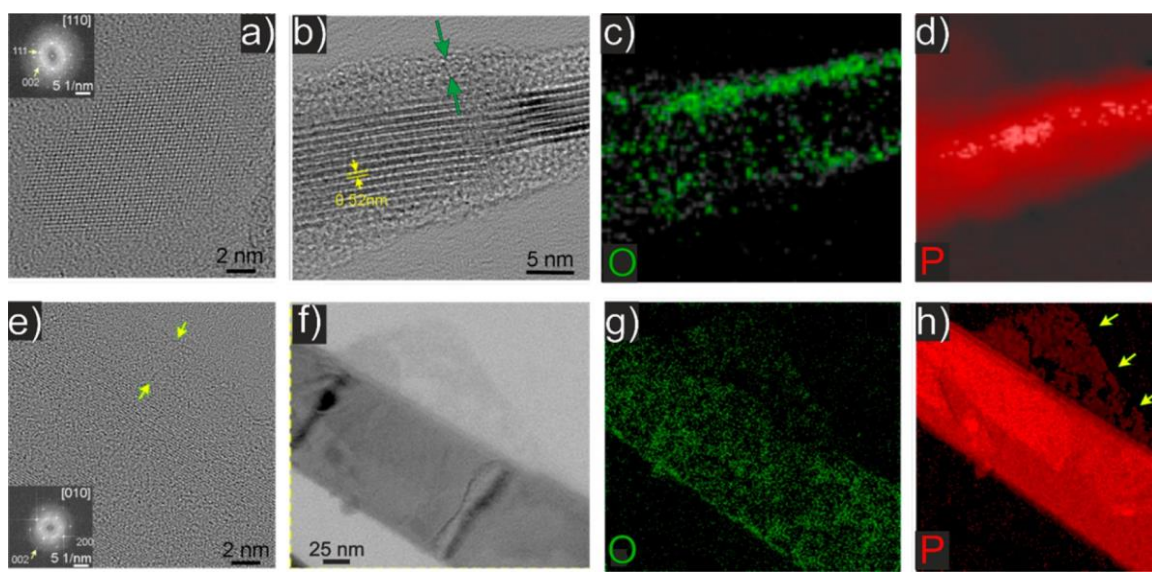

*Figure 4*. Study on the structural degradation of (a–d) FL-BPacetone and (e–h) FL-BPCHP flakes after 12 weeks of storage in the respective solvents. (a) An FL-BPacetone flake in the 110 zone axis exhibiting the structure of bulk BP. The corresponding FT is shown as an inset to part a. (b) An 8layer-thick flake shown edge-on, in which the individual phosphorene monolayers of ∼0.52 nm in

*thickness within the flake can be clearly resolved. The same flake also exhibits an amorphous layer on the surface (see green arrow), and the corresponding (c) oxygen and (d) phosphorus EFTEM elemental maps indicate that this layer is oxygen-rich. A thick amorphous layer is also present on the FL-BP$_{CHP}$ flakes (see arrows in part e). The flake shown in part e is in the 010 zone axis, with the corresponding FT shown as an inset. Additionally, a thinner (brighter in contrast in the TEM image in part f) flake overlapping the flake shown in part e exhibits clear signs of structural degradation. The corresponding (g) oxygen and in particular (h) phosphorus EFTEM elemental maps clearly indicate the fragmentation of the thinner flake (yellow arrows in part h).*

### Application of FL-BP Flakes as an Active Material for the LIB Anode.

The successful exfoliation of BP in acetone exploiting a low-boiling-point solvent, not toxic, not degrading with the annealing temperature, offers the possibility to scale up the production,[1,28,29,35] for applications in the energy storage sector. To further highlight this, below we compare the electrochemical performances of FL-BP$_{acetone}$ for the fabrication of electrodes for Li-ion batteries, with those of FL-BP$_{CHP}$. To this end, both dispersions are mixed with a conductive agent and a binding material, *i.e.*, CB and PVdF, respectively, then deposited onto copper substrates (see experimental section for details). The SEM images of the samples show the mixture of the FL-BP, PVdF, and CB covering the copper substrates (see Figure 5a,b, obtained from acetone and CHP dispersions, respectively). The optical pictures, reported as insets in Figure 5a,b, show the copper substrate coated with the FL-BP/CB/PVdF. These images demonstrate that a complete and uniform coverage of the substrate is achieved with the FL-BP$_{acetone}$ sample. In contrast, for the deposited FL-BP$_{CHP}$, the substrate is not uniformly coated. This inhomogeneous material distribution is attributed to the slow drying/evaporation of the CHP (hours time scale), and the temperature required to dry the electrode (~180 °C).

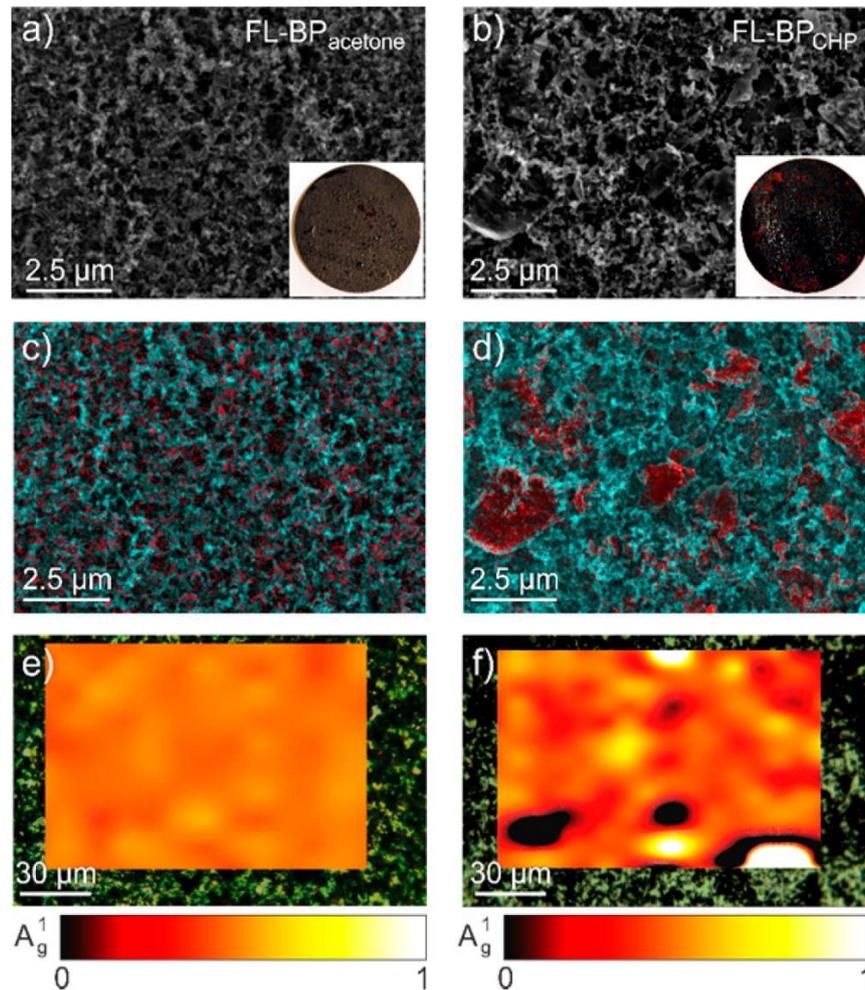

*Figure 5. Scanning electron microscopy images of electrodes made of (a) FL-BPacetone and (b) FL-BPCHP, mixed with CB and PVdF deposited onto copper substrates. Energy-dispersive X-ray spectrometry elemental mapping of carbon (cyan) and phosphorus (red) of (c) FL-BPacetone and (d) FL-BPCHP. Raman mapping on the electrodes plotting the signal to baseline intensity of the A1g peak of (e) FLBPacetone and (f) FL-BPCHP.*

The heterogeneity in terms of material distribution, *i.e.*, FL-BP and CB/PVdF, on the electrodes is also analysed by using EDS and Raman mapping. The presence of FL-BP is characterized by means of EDS, analysing the Kα energy of phosphorus (2.013 eV), while the presence of CB or PVdF is identified by investigating the Kα energy of carbon (0.277 eV). In the FLBP$_{acetone}$ sample, the EDS mapping (Figure 5c) shows that the phosphorus, in red, is homogeneously distributed onto the substrate and evenly spread with the carbon, in cyan. On the contrary, for the FL-BP$_{CHP}$ sample (Figure 5d), the phosphorus distribution suggests that the FL-BP is aggregated, while the carbon is uniformly distributed. The Raman mapping of the film like deposited samples further supports the different distribution of FL-BP in the two samples by monitoring the signal to baseline ratio of the A$^1_g$ peak. The Raman mappings of FL-BP$_{acetone}$ and FL-BP$_{CHP}$ are shown in Figure 6e,f, respectively. The Raman mapping of FL-BP$_{acetone}$ shows the homogeneous presence of the A$^1_g$ peak, with an average intensity of 0.5, which confirms the uniform distribution of FLBP crystals onto the substrate. On the contrary, the FL-BP$_{CHP}$ Raman mapping presents preferential areas where the A$^1_g$ peak is concentrated (intensity 1) and others where the A$^1_g$ peak is not present (intensity 0), indicating aggregation of the FL-BP flakes.

The FL-BP-based anodes are then tested against a Li foil in a half-cell configuration, as described in the methods section. The electrochemical results of the FL-BP anodes are summarized in Figure 6. The cyclic voltammetries (CVs; Figure 6 a) are performed at a rate of 30 µV s$^{-1}$ to obtain an electrochemical response for the Li ion transfer from the Li foil to the FL-BP based anodes.[90,93] The CV scan ranges from 0.05 to 3 V versus Li$^+$/Li, which is within the reaction range of both the formation of the solid–electrolyte interface (SEI) and the lithiation/ delithiation processes for the BP material.[14] The first CV reduction cycle shows multiple peaks around 0.6–1.0 V, which are attributed to the phase change from BP → Li$_x$P → LiP → Li$_2$P → Li$_3$P.[14]

The voltage profiles of the FL-BP electrodes during the 1st and 20th galvanostatic charge/discharge cycles are performed at a specific current of 100 mA g$^{-1}$ between 50 mV and 3 V versus Li$^+$/Li, to complete the lithiation/delithiation (charge/ discharge) process during each cycle (Figure 6b). From the first voltage profile of the FL-BP$_{acetone}$ anode, an initial capacity of 1732 mA h g$^{-1}$ and a discharge capacity of ∼510 mA h g$^{-1}$ are measured. Such a high irreversible capacity (∼1220 mA h g$^{-1}$) is typical of nano flake size-based anodes,[12,91,92] and similar behaviour has already been reported for BP-based anodes.[14,54] The high irreversible capacity is associated with the large quantity of Li ions that are consumed for the SEI formation on the FL-BP large surface area and trapped by the high-energy binding on the edges.[12,91,92] The voltage profiles of both FL-BP anodes show that more than 80% of the electrode capacity is delivered at a potential that is lower than 1 V versus Li$^+$/Li over the 20 cycles. Such a low potential is beneficial for the application of the FL-BP anode material to target high-energy efficiency Li-ion batteries.[12]

As shown in Figure 6c, both FL-BP anodes present a significant capacity which fades within the first 10 cycles. This could be caused by the large volume change that originates from the lithiation/delithiation processes during different Li$_x$P phases.[14] However, the FL-BP$_{acetone}$ anode stabilizes at a specific capacity of ∼480 mA h g$^{-1}$, with a Coulombic efficiency of 99.6% after 100 charge/discharge cycles, taken at a current density of 0.1 A g$^{-1}$. Meanwhile, the anode based on FL-BP$_{CHP}$, tested under the same experimental conditions, shows a specific capacity of ∼200 mA h g$^{-1}$ with a Coulombic efficiency of 99.6%. For further insights on the performances, the two FL-BP electrodes are cycled between 50 mV and 3 V versus Li$^+$/Li at specific currents ranging from 0.1 to 1 A g$^{-1}$ to investigate the electrochemistry activities of the samples during fast lithiation/delithiation (charge/discharge) processes. The results, presented in Figure 6d, demonstrate that the FL-BP$_{acetone}$ electrode presents stable discharge cyclability with a specific capacity of 447 mA h g$^{-1}$ at a specific current of 0.2 A g$^{-1}$ after the 20th charge/discharge and a Coulombic efficiency of 99.7%. On the other hand, under the same experimental conditions, the FLBP$_{CHP}$ electrode reaches a specific capacity of 185 mA h g$^{-1}$ and a Coulombic efficiency of 99.4% (Figure 6c,d).

Furthermore, for the FL-BP$_{acetone}$-based anode, specific capacities of 382 and 345 mA h g$^{-1}$ are reached at current densities of 0.5 and 1 A g$^{-1}$, respectively. Only less than a 30% drop in the specific capacities (from 480 mA g$^{-1}$ to 345 mA h g$^{-1}$) is observed for the FL-BP$_{acetone}$ anode which was tested at both a low (0.1 A g$^{-1}$) and a high (1 A g$^{-1}$) current density, indicating that the FL-BP$_{acetone}$-based battery is a promising option for fast charge/discharge devices.[16,93,94] Although the specific capacity of the FL-BP$_{CHP}$ anode also presents a drop of 10% with the specific current varying from 0.1 to 1 A g$^{-1}$, its specific capacity still remains below 200 mA h g$^{-1}$, i.e., 50% lower than the one based on FL-BP$_{acetone}$. Moreover, the FLBP$_{acetone}$-based anode outperforms previously reported anodes based on solution-processed BP (i.e., ~200 mA h g$^{-1}$ at 0.1 A g$^{-1}$ after the second discharge cycle,[45] and ~250 mA h g$^{-1}$ at 0.1 A g$^{-1}$ after the first discharge cycle[54]). We attribute the difference in the electrochemical performances of the two different anodes, i.e., the specific capacity, to the aggregation of the FL-BP$_{CHP}$ after the deposition, as demonstrated by both the SEM and Raman mappings reported in Figure 5. In contrast, the FL-BP$_{acetone}$ flakes do not aggregate after the deposition onto the copper substrate.

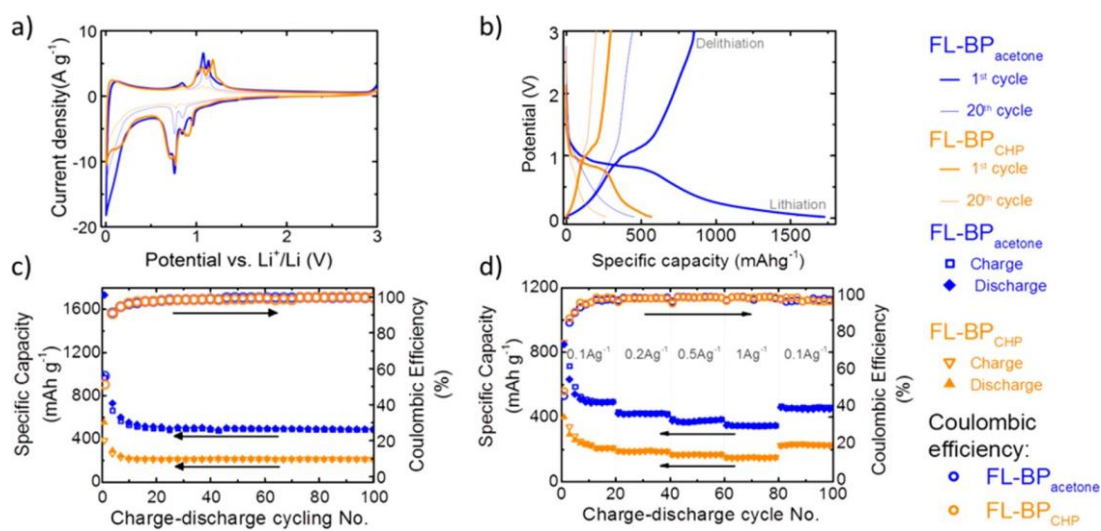

*Figure 6*. *Electrochemical analysis of the FL-BP film (acetone-exfoliated sample data points shown in blue; CHP-exfoliated sample data points shown in orange). (a) Cyclic voltammograms at a scan rate of 30 μV s–1 and (b) voltage profile upon galvanostatic charge/discharge of BP-based electrodes at 0.1 A g–1 between 50 mV and 3 V. Thick and thin lines correspond to 1st and 20th cycle, respectively. Specific capacity and Coulombic efficiency over galvanostatic charge/discharge and cycling (c) between 50 mV and 3 V at 0.1 A g–1 and (d) different specific currents, for both the FL-BPacetone- and FL-BPCHP-based electrodes.*

## Results and discussion

In conclusion, we demonstrated the exfoliation of bulk black phosphorus with acetone, creating a new way to formulate functional inks to be exploited for the designing of few-layer BP-based devices. The exfoliation of BP and its stable dispersion in 14 different solvents give us the possibility to evaluate the dispersibility properties of few-layer BP. We have indeed estimated the following: (1) the surface energy to be in the range 55–70 mJ m$^{-2}$; (2) the Hildebrand parameter to be 21 MPa$^{1/2}$; and (3) the Hansen parameters to be 15–18 MPa$^{1/2}$ for a dispersive force, 5–12 MPa$^{1/2}$ for a polar force, and 5–10 MPa$^{1/2}$ for a hydrogen bonding force. Among the different trials, we found that exfoliation in acetone leads to exfoliated FL-BP flakes with average lateral size of 30 nm and thickness of 7 nm (corresponding to 13 layers). Additionally, we demonstrated, by electron energy loss and Raman spectroscopies, that the exfoliated BP flakes in acetone show an aging (oxidation and degradation) that is comparable with the one obtained by using high-boiling-point solvents, in specific cyclohexyl-2-pyrrolidone. Having been motivated by these results, we exploited BP flakes which were exfoliated in acetone as an active material for the realization of LIB anodes. We have seen that FL-BP$_{acetone}$ flakes are homogeneously distributed onto the current collector substrate thanks to the fast solvent removal. The FL-BP$_{acetone}$-based anodes, tested in half-cell configuration, achieved a specific capacity of 480 mA g$^{-1}$ at a current density of 0.1 A g$^{-1}$, with a Coulombic efficiency of 99.6% after 100 charge/discharge cycles. The FLBP$_{acetone}$-based anode outperformed the FL-BP$_{CHP}$-based one (∼200 mA h g$^{-1}$ after 100 cycles at 0.1 A g$^{-1}$, achieving a Coulombic efficiency of ∼99.0%). The proposed liquid-phase exfoliation process can also be scaled-up since the use of acetone does not present environmental risks, whereas both CHP and NMP do. Finally, we can further improve the exfoliation of BP in acetone by the addition of acetone-soluble polymers, thus enabling the large-scale production of FL-BP/ polymer composites.

## Acknowledgements

The authors acknowledge the European Union's Horizon 2020 research and innovation programme under Grant Agreement 696656⎯GrapheneCore1 and the Seventh Framework Programme under Grant Agreement 614897 (ERC Consolidator Grant "TRANS-NANO") for financial support.

# Appendix

## Exfoliation process.

The exfoliation of Black phosphorus (BP) is analysed by using 14 different solvents. The solvents with different characteristics: surface tension, Hildebrand and Hansen parameters, toxicity, and boiling points, are listed in table S1.

*Table S1*. List of solvents with their corresponding surface tension, Hildebrand and Hansen parameters, and boiling points.

| Solvent | Surface Tension (mNm$^{-1}$) | Hildebrand parameter (MPa$^{1/2}$) | Hansen parameter. Dispersive force (MPa$^{1/2}$) | Hansen parameter. Polar force (MPa$^{1/2}$) | Hansen parameter. Hydrogen bonding force (MPa$^{1/2}$) | Boiling point (°C) |
|---|---|---|---|---|---|---|
| Acetone | 22.2 | 19.9 | 15.5 | 10.4 | 7.0 | 56.0 |
| Toluene | 28.4 | 18.2 | 18.0 | 1.4 | 2.0 | 110.6 |
| Chloroform | 25.8 | 18.9 | 17.8 | 3.1 | 5.7 | 61.2 |
| 2-Propanol | 20.6 | 23.6 | 15.8 | 6.1 | 16.4 | 82.6 |
| Trichloroethylene | 28.7 | 19.0 | 18.0 | 3.1 | 5.3 | 87.2 |
| Methanol | 21.8 | 29.8 | 15.1 | 12.3 | 22.5 | 64.7 |
| Ethylene glycol | 47.0 | 33.0 | 17.0 | 11.0 | 26.0 | 197.3 |
| Acetonitrile | 27.7 | 24.4 | 15.3 | 18.0 | 6.1 | 82.0 |
| Ethanol | 21.1 | 26.5 | 15.8 | 8.8 | 19.4 | 78.4 |
| n-Hexane | 18.7 | 14.9 | 14.9 | 0.0 | 0.0 | 68.0 |
| N-Methyl-2-pyrrolidone | 40.1 | 23.0 | 18.0 | 12.3 | 7.2 | 202.0 |
| Dimethylformamide | 37.1 | 24.9 | 17.4 | 13.7 | 11.3 | 153.0 |
| Diethyl carbonate | 28.1 | 18.7 | 15.5 | 3.9 | 9.7 | 144.7 |
| N-Cyclohexyl-2-pyrrolidone | 43.2 | 20.5 | 18.2 | 6.8 | 6.5 | 284.0 |

The approach of "like dissolves like" has been used previously to have an estimation of the solubility properties of nanotubes[1,2,3], graphene[4,5], carbon fibres[6], and 2D crystals[7]. The process consists in suspending the grinded BP in a solvent and applies ultrasonic waves to promote the exfoliation. The exfoliation temperature plays a crucial role, especially with the low boiling point solvents (see Table S1), because of the solubility parameters changes *e.g*, surface tension. The sonic bath temperature during the exfoliation process is controlled to be in the 35-40 °C range. After the centrifugation/decantation the extinction spectrum are taken, Figure S1.

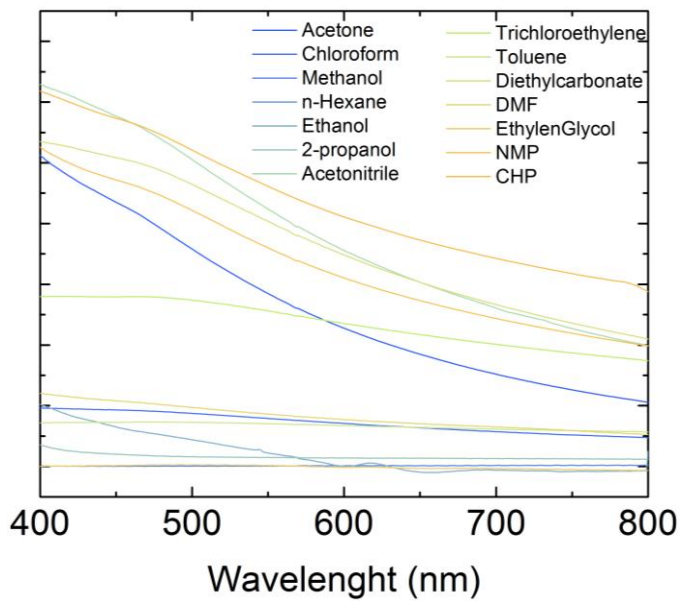

Figure S1. *Extinction spectra of BP after LPE.*

By taking the value of the extinction at 680 nm of the dispersion in different solvents, and plotting them against the surface tension of each solvent, it is possible to estimate the surface tension of the suspended material. This approach is widely used in polymers science to determine the solubility properties of a polymer.[8,9] We applied the same approach in the figure 1b and 1c in the main text, the plot 1b indicates that the solvents with surface tension in the range between 30 to 50 mNm$^{-1}$ are the solvents with the required surface tension to exfoliate and stably disperse few layers (FL) BP. The extinction values from the dispersions in different solvents give us also information about the surface energy of exfoliated BP, by using the approximation $\gamma = E_{Surface}^{Solvent} - TS_{Surface}^{Solvent}$, where $\gamma$ is the surface tension, *E* is the solvent surface energy, *T* is the absolute temperature and *S* is the solvent surface entropy, which generally takes a value of 0.0001 J m$^{-2}$ K$^{-1}$.[6,10] Solving for *E*, the surface energy of BP is found to be in the range between 55 and 70 mJ m$^{-2}$.

In the same manner, the Hansen parameters of the BP are estimated (Figure S2), indicating a dispersive force: 15-18 MPa$^{1/2}$; a polar force: 5-12 MPa$^{1/2}$ and hydrogen bonding force: 5-10 MPa$^{1/2}$. This means that a solvent with these Hansen parameters should be able to exfoliate and suspend the FL-BP.

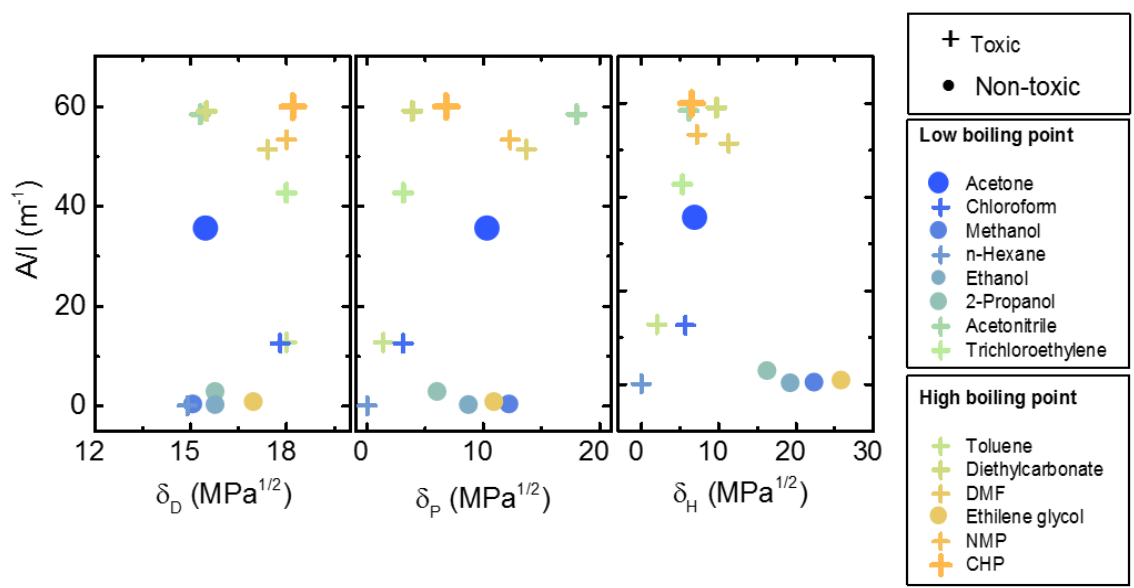

*Figure S2*. Estimation of the Hansen parameters of FL-BP, giving approximated parameters of dispersive force: 15-18 MPa1/2; polar force: 5-12 MPa1/2 and hydrogen bonding force: 5-10 MPa1/2

### Transmission Electron Microscopy

In order to assess the ageing behaviour of the FL-BP exfoliated and stored in N-cyclohexyl-2-pyrrolidone (CHP) and in acetone, a set of imaging and analytical TEM techniques are used to periodically monitor the oxidation and structural degradation of the FL-BP flakes.

The comparison of the electron energy loss (EEL) spectra collected from the flakes after different storage times in CHP (Fig S3 a)) and in acetone (Fig S3 b)) indicates that both samples undergo gradual oxidation over a three months period following the exfoliation. This is evident from the rise of the peak at ~136 eV, in addition to the P $L_{2,3}$-edge at ~130 eV (labelled as $P^0$) which corresponds to elemental P (Figure S3 a) and b)). The peak at ~136 eV (labelled as $P_XO_Y$) has previously been attributed to oxidation of BP flakes [11,12,13,14] and is also a dominant feature in the P $L_{2,3}$-edge from $P_2O_5$.[11,12,13] The corresponding EEL spectra focused on the oxygen K-edge region (Fig S3 c)) from the two samples following more than three months of ageing, exhibit weak but distinguishable signals associated with the presence of oxygen (arrowed). In contrast, no clear signal in the same energy range is observed in the spectra from the as-exfoliated samples.

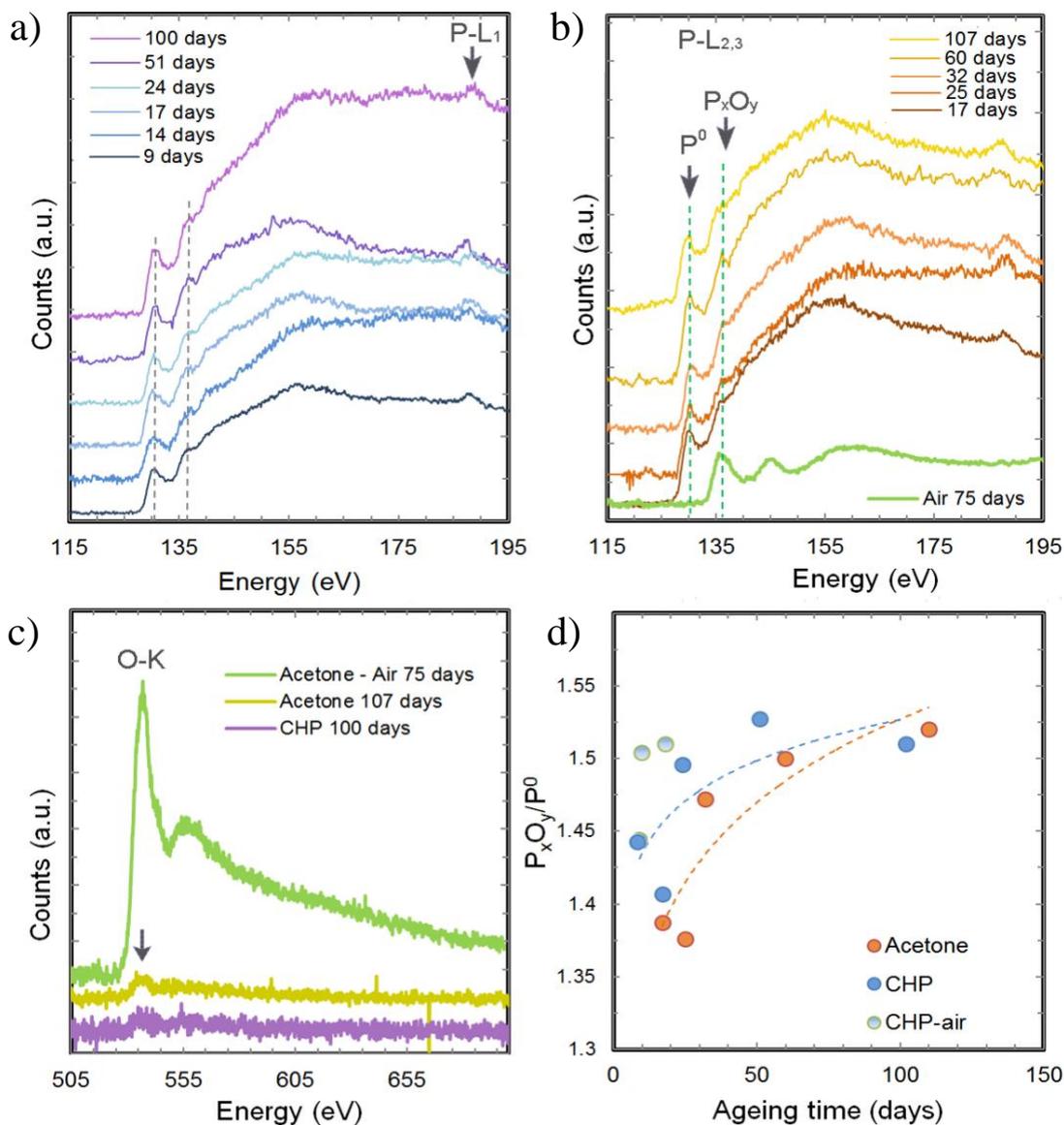

*Figure S3.* The EELS characterization of the FL-BP degradation in acetone and in CHP. EEL spectra exhibiting the phosphorus L-edge, collected from the FL-BP flakes stored in (a) CHP and in (b) acetone for different time, indicated in the graphs, following the exfoliation in the respective solvents. An EEL spectrum from the flakes stored in air for 75 days following the exfoliation in acetone is included for comparison in (b). The P L2,3-edge at ~130 eV corresponds to elemental P (P0), while the appearance of this edge at higher energies indicates that P is in a higher oxidation state. The additional peak observed here at ~136 eV has been attributed to oxidation of FL-BP flakes,11,12,13,14 hence it is labelled PXOY. (c) EEL spectra in the oxygen K-edge region collected from the FL-BP flakes stored in CHP and in acetone for 100 and 107 days, respectively. The oxygen K-edge signal from the flakes after storage in the respective solvents is considerably weaker than the oxygen signal from the acetone-exfoliated flakes stored in air for 75 days, included for

*comparison in (c). (d) Evolution of the intensity ratio of the PXOY to P0 signals in the EEL spectra collected after different storage times in CHP (blue circles) and in acetone (orange circles). Additional data from the CHP-exfoliated sample stored in air up to two weeks are also included for comparison in (d) (open blue circles). Each data point is an average of six measurements. The dashes lines are power-low fits to the experimental data points.*

Although no significant differences are immediately visible between the EEL spectra from the two samples after comparable storage times in their respective solvents, the intensity ratio of the $P_xO_y$ signal at 136 eV versus the $P^0$ signal at 130 eV, plotted in Figure S3 (d), is slightly but consistently higher in the spectra collected from the FL-BP $_{CHP}$ over the initial two months. After more than three months of storage (~100 days of storage in the respective solvents), both samples reach a similar level of oxidation (1.52 $P_xO_y$ to $P^0$ ratio, Fig S3 d). The rate of oxidation of the FL-BP in both solvents is nevertheless significantly lower than in air (additional data points from the FL-BP$_{CHP}$ left in air for two weeks are included for comparison; light blue circles in Fig S3 d). This is evident from the prominent oxygen K-edge in the spectrum from the acetone-exfoliated sample left in air for 75 days (the topmost spectrum in Fig S3 c) compared to the oxygen signals from flakes stored in the two solvent for more than three months.

The FL-BP flakes exposed to air in the present work retained their crystalline structure and most of the phosphorous in the elemental form within the initial couple of weeks, however the prolonged exposure to air gradually lead to their oxidation and structural degradation. The EEL spectrum from the acetone-exfoliated sample left in air for 75 days included in Fig S3b) (green line) indicates a chemical shift from 130 eV to ~136 eV and exhibits additional features at higher energies, both consistent with the formation of phosphorus oxide.[11,12,13,14] The oxidation is accompanied by a significant structural transformation. What are initially fine crystalline FL-BP flakes in aggregates (Figure S4 a)), after exposure to air for 75 days transformed into amorphous clusters (Figure S4 b)), similar to droplet-like features reported previously.[13,14] The compositional analysis based on EELS indicates that the composition of the amorphous clusters is approximately $P_{54}O_{46}$ (expressed in at%; Figure S4 c). The compositional analysis is based on the P L-edge found here at 136 eV and O K-edge at ~532 eV, both extracted from the raw spectra by removing the background fitted according to a power law model and using the Hartree-Slater model for the cross-section calculation.

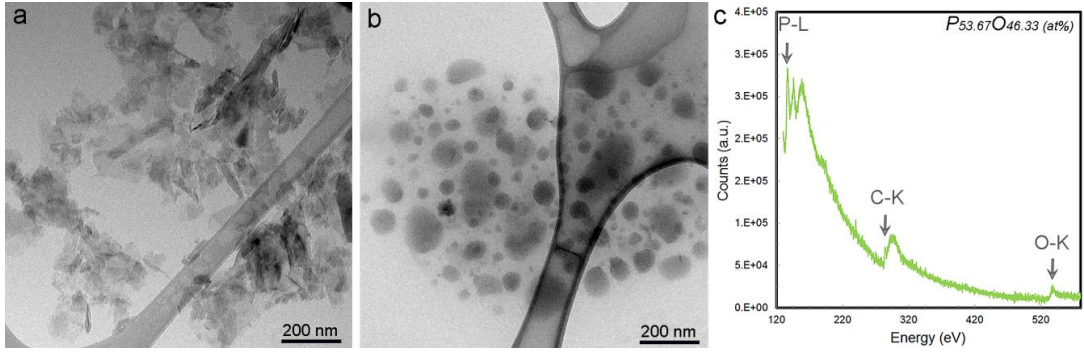

*Figure S4*. *The degradation of the FL-BP flakes in air. (a) An aggregate of flakes after exfoliation in acetone and (b) a similar aggregate after exposure to air for 75 days. (c) An EEL spectrum from the flakes after oxidation in air for 75 days exhibiting prominent phosphorus and oxygen edges used for quantification. The carbon K-edge at 284 eV originates from the amorphous carbon TEM grid support.*

## Sedimentation based separation, the solvent viscosity and density importance on the flake morphology.

In the main text, we stated that the forces interacting on a particle when is subjected at a force field are the addition of the centrifugal force ($F_c$), the buoyant force ($F_b$) and the frictional force ($F_f$), represented as:

(Eq. S1) $$F_c - F_b - F_f = F_{tot}$$

Where, $F_c = m_p \omega^2 r$, and $m_p$ is mass of the particle, $r$ is the distance from the rotational axes, and $\omega$ the angular velocity. $F_b = -m_s \omega^2 r$, $m_s$ is the mass of the displaced solvent, $F_f = -fv$, $v$ is the sedimentation velocity and $f$ is the friction coefficient between the solvent and the particle. If we consider that the volume of the particle or flake can be caculated as its mass ($m_p$) divided by its density ($\rho_p$), the displaced solvent can be calculated as $m_s = m_p \rho_s/\rho_p$, where $\rho_s$ is the density of the solvent. The expression substituting each force in the Eq. S1 gives:

(Eq. S2) $$m_p \omega^2 r - m_p \omega^2 r \frac{\rho_s}{\rho_p} - fv = m_p \frac{dv}{dt}$$

When the flakes are in equilibrium the *dv/dt* = 0, specifically, the sum of the forces acting on the flakes dispersed in the solvent is zero. The Eq. S2 can be re- written as:

(Eq. S3) $$S = \frac{v}{\omega^2 r} = \frac{m_p\left(1-\frac{\rho_s}{\rho_p}\right)}{f}$$

At this point, we can then define the ratio of sedimentation, or sedimentation coefficient (*S*) as the ratio between the sedimentation velocity and the particle acceleration, *i.e.* the centrifugal acceleration ($\omega^2 r$) in this case. From this equation we can derive the equation 5 in the main text.[15]

If we model our flakes as a spheroid, it is possible to approximate the particle mass as $m = \frac{4\pi}{3} ab^2 \rho_p$, where $a$ and $b$ are the minor and the major semi-axes of the ellipsoid, respectively, see the figure S5, and $\frac{4\pi}{3} ab^2$ is its volume. The $f$ can be computed as[16]

(Eq. S4) $$f = F_t f_0$$

where $f_0$ is the friction coefficient of a sphere having the same volume of the investigated particle and $F_t$ is the frictional ratio, which is a the geometrical correction factor for the actual shape of the given particle. Assumed that the friction factor for a sphere is $f = 6\pi\eta r$, we can find that:

(Eq. S5) $$f_0 = 6\pi\eta R_e$$

where $\eta$ is the solvent viscosity and the equivalent radius $R_e$ of the ellipsoid is:

(Eq. S6) $$R_e = \sqrt[3]{ab^2}$$

$F_t$ for an spheroid is purely geometric and is calculated as:[17]

(Eq. S7) $$F_t = \frac{\sqrt{\left(\frac{b}{a}\right)^2 - 1}}{\left(\frac{b}{a}\right)^{\frac{2}{3}} arctan\sqrt{\left(\frac{b}{a}\right)^2 - 1}}$$

By combining Eq. S4 and S5 and using the results of Eq. S7 we obtain:

(Eq. S8) $$f = 6\pi\eta R_e F_t$$

Following Eq. 5 and S8, the $S$ decreases with the viscosity and solvent densiy, while increases according to the mass of the flake and decreases with its size. Table S2 shows the S for the samples FL-BP$_{CHP}$ and FL-BP$_{Acetone}$ under the size constrains given by the microscopy techniques (in blak letters), and, as demonstration, under the condition where the flake dimensions are the same, but the solvens are different. The $S$ ratio, for the same size flakes (4.05 nm in thickness radio and 30 equivalent radio) in different solvents, *i.e.* 5.59×10$^{-13}$ s and 1.73×10$^{-11}$ s for FL-BP$_{acetone}$ and FL-BP$_{CHP}$ respectively, is 31. This means that flakes dispersed in acetone will precipitate 31 times faster than in CHP, when are subjected to a centrifugal force.

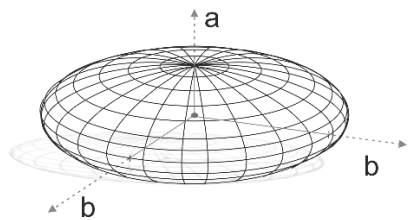

*FigureS4.* Schematic representation of a flake considering it as a spheroid.

*Table S2*. Considered parameters to calculate the sedimentation coefficient for the exfoliated BP flakes, considering a rotor radius 150 mm, and 900 rpm. In black letters the real size of the flakes measured by microscopy techniques, in grey proposed flakes sizes to obtain the sedimentation ratio for same-size flakes in different solvents.

|  | Thickness, (nm) | (Lateral size)/2 (nm) | Frictional ratio | Viscosity | Friction coefficient | Partice density | Solvent density | Mass | Sedimentation coefficient |
|---|---|---|---|---|---|---|---|---|---|
|  | $a$ | $b$ | $F_t$ | $\eta$ (mPa s) | $f$ | $\rho_p$ (g/cm$^3$) | $\rho_s$ (g/cm$^3$) | $m_p$ (kg) | $S$ (s) |
| FL-BP$_{acetone}$ | 4.05 | 30.0 | 1.35 | 11.5 | 2.24×10$^{-9}$ | 2.29 | 0.78 | 1.91×10$^{-21}$ | 5.59×10$^{-13}$ |
| FL-BP$_{CHP}$ | 4.05 | 30.0 | 1.35 | 0.316 | 6.17×10$^{-11}$ | 2.29 | 1.01 | 1.91×10$^{-21}$ | 1.73×10$^{-11}$ |
| FL-BP$_{acetone}$ | 3.50 | 15.0 | 1.18 | 11.5 | 1.18×10$^{-09}$ | 2.29 | 0.78 | 4.12×10$^{-22}$ | 2.29×10$^{-13}$ |
| FL-BP$_{CHP}$ | 3.50 | 15.0 | 1.18 | 0.316 | 3.25×10$^{-11}$ | 2.29 | 1.01 | 4.12×10$^{-22}$ | 7.10×10$^{-12}$ |